# Linear and Nonlinear Optical Properties of Mn doped Benzimidazole Thin Films

P.A. Praveen[1], R. Ramesh Babu[1,*], SP. Prabhakaran[1], K. Ramamurthi[2]

[1]*Crystal Growth and Thin film Laboratory, Department of Physics, Bharathidasan University, Tiruchirappalli-620024, Tamilnadu, India.*
[2]*Department of Physics and Nanotechnology, Faculty of Engineering and Technology, SRM University, Kattankulathur – 603 203, Tamil Nadu, India.*
*\*rampap2k@yahoo.co.in*

**Abstract.** In the present work, the Mn doped benzimidazole (BMZ) thin films were prepared by simple chemical bath deposition technique. The material was directly deposited as thin film on glass substrates and the metal concentration in the solution was varied in weight percentage in order to investigate the dopant effect on the properties of thin films. Similarly, the Mn doped BMZ films were deposited in different solution temperature to study the effect of deposition temperature on the properties of thin films. The PXRD and FT-IR spectroscopy are used to study the structural and the presence of functional groups in the BMZ medium. Depending upon the solution temperature, thickness of the films varying from 0.6 to 1.2 μm and the optical transparency of the samples increases with the increasing temperature up to 50 °C. Second Harmonic Generation (SHG) efficiency of the films is measured for all the films. Third order nonlinear optical properties of the films were analyzed using Z-scan technique. The experimental results show that Mn doped BMZ films exhibits saturation absorption and negative nonlinearity.

## INTRODUCTION

Metal doped organic materials are the better candidates for photonic applications. These type of materials can largely contribute for Third Harmonic Generation (THG) due to their large polarizabilities and charge transfer mechanism. BMZ is a well known organic NLO material and the studies on the effect of metal ions in this system will be fruitful [1]. In the present work, Mn doped BMZ films were prepared by simple chemical bath deposition method with different solution temperatures. The variation in optical transmittance, Second Harmonic Generation (SHG) efficiency, third order nonlinear susceptibility values with respect to the solution temperature and Mn ion concentration in the solution is studied.

## EXPERIMENTAL

All the reagents used for synthesis are commercially available and purchased from Merck chemicals. Microscopic glass slides (Labtech) were used as the substrates for the films deposition. Ligand solution was prepared by dissolving BMZ (0.236 g, 0.1 M) in ethanol (20 ml) and the dopant [$(CH_3COO)_2$Mn.$4H_2O$] was added to the solution in different weight percentages 1.0, 3.0, 5.0, 7.5 and 10 wt%. Two glass substrates were fused and placed inside the solution. Then the solution was kept in a temperature bath for an hour and the solution temperature was varied by adjusting the bath temperature. Films were prepared at different solution temperatures. Then the substrates with film was removed from the solution container and dried in a hot air oven at 50 °C for 10 minutes.

The deposited films were subjected to various characterization studies in order to explore the effect of metal ions concentration and solution temperature. Reflection and transmission spectra of the samples were recorded using Perkin Elmer Lambda 35 UV-VIS spectrometer. SHG efficiency of the films was analyzed using a Nd:YAG laser of wavelength 1064 nm and the output from the samples was measured at 532 nm using a power meter. Third order nonlinear optical properties of the films were studied by using Z-Scan technique at room temperature. A CW diode laser of wavelength 650 nm and 5 mW power was used as the source. The beam of intensity about $7.8 \times 10^8$ W/cm$^2$ was focused on to the sample using a lens of focal length 15 cm and scanned along the Z-axis with respect to the direction of propagation of the laser beam. Two types of configuration are possible by adjusting the aperture linear transmittance *S*. When *S*=1, configuration called open aperture (OA), the detector was insensitive to the nonlinear refractive (NLR) effects and nonlinear absorptive (NLA) effects could be evaluated. In order to study the refractive effects, the second configuration; closed aperture, with 40% of transmittance (*S*=0.4) was used and it was divided by OA data to ensure the presence of complete NLR effects only.

## RESULTS AND DISCUSSION

### Structural Analysis

The recorded diffraction pattern for 5 wt% of Mn doped BMZ films at 50 °C is given in the Figure 1. The pattern shows two intense peaks corresponding to (002) and (004) planes and a peak with less intensity along the (110) plane. (110) plane is the dominant peak for BMZ [2] whereas in the present case (002) and (004) planes. This variation can be attributed to the effect of Mn ions which may occupy in the interstitial sites of the BMZ medium and may creates a lattice mismatching in the crystalline film leads to the observed pattern. The variation of solution temperatures does not show any significant changes with the XRD pattern, but intensity of (002) and (004) planes increases with increasing dopant concentration.

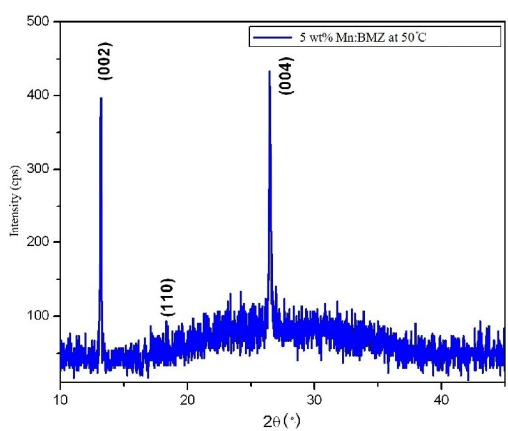

**FIGURE 1.** Powder XRD patterns of Mn doped BMZ films deposited for 5 wt % at 50 °C.

The FT-IR spectrum of Mn doped BMZ films were recorded between 400 and 4000 $cm^{-1}$ and the spectrum was compared with already available report on BMZ crystal [2]. Generally, the heterocyclic aromatic compounds may coordinate with Mn through N in the imidazole ring. The peak at 530 $cm^{-1}$ is attributed to the presence Mn ions in the film [3]. This peak shifted towards the lower wavenumber region with increasing dopant concentration (Table 1). It was observed that increasing the dopant concentration increases the absorbance in the region of 530 $cm^{-1}$. This is due to increase in number of Mn ions bonding with N in the imidazole ring [3].

**TABLE 1.** Observed peak shift ($cm^{-1}$) of Mn:BMZ* films

| 1 wt% | 3 wt% | 5 wt% | 7.5 wt% | 10 wt % |
|---|---|---|---|---|
| 538 | 532 | 526 | 514 | 501 |
| #T% | T% | T% | T% | T% |
| 6.96 | 6.50 | 6.38 | 5.33 | 5.82 |

*Reported value 545 ($cm^{-1}$) [3]
# T% - Transmittance percentage

### Linear Optical Properties

The transmittance spectrum of Mn doped BMZ thin films was recorded in the wavelength range from 300 to 1100 nm and is shown in the Figure 2. In the present work, increasing dopant concentration upto 5 wt% results good transparency to the films. Further, increasing solution temperature also leads to good optical transparency. The reason is temperature may provide necessary energy for the BMZ molecules to adhere on the substrate. At temperatures higher than 50 ˚C, evaporation of ethanol increases leads to the rapid adsorption of the material, which in turn reduces the optical transparency. Therefore, 5 wt% of Mn in BMZ solution at 50 ˚C are chosen as the optimum condition to deposit Mn:BMZ thin films.

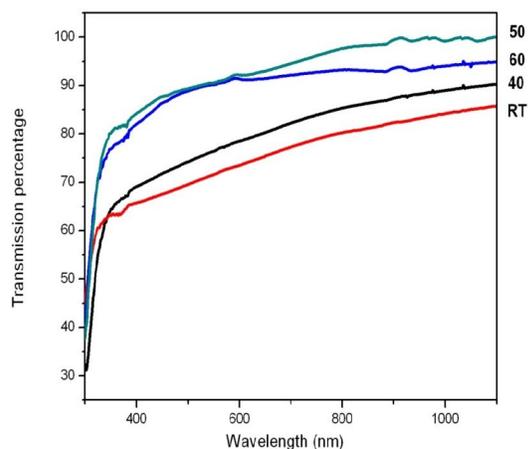

**FIGURE 2**. Linear optical transmittance of Mn :BMZ films for different solution temperature

## SHG Efficiency

The second harmonic generation of the films were confirmed by subjecting the films in front of Nd:YAG laser of wavelength 1064 nm. The second harmonic efficiency of the deposited films was compared with the pure BMZ films. The corresponding values are given Table 2. 5 wt% Mn:BMZ films deposited at 50°C show higher emissivity. The increase in SHG efficiency can be attributed to the homogeneity of the films which is in analogues to transmission spectrum.

**TABLE 2.** SHG efficiency of 5 wt% Mn:BMZ samples

| Sample Prepared at (°C) | SHG Efficiency* |
|---|---|
| RT | 1.5714 |
| 40 | 1.6285 |
| 50 | 1.7140 |
| 60 | 1.6857 |

* Compared with SHG of pure BMZ films

## Third Harmonic Susceptibility analysis

The third-order nonlinear refractive index and the nonlinear absorption coefficients were evaluated for Mn doped films for optimized experimental conditions (5 wt% Mn:BMZ at 50 °C) by the Z-scan technique. Figure 3 shows the open and closed aperture Z-scan curves of the Mn doped BMZ films. The obtained open aperture curve is symmetric with respect to focus (Z=0) where it has minimum transmittance for nonlinear absorption due to reverse saturable absorption (RSA). The RSA behavior of Mn doped BMZ thin films can be inferred with the help of five-level model [4]. It is due to weak ground state absorption between $S_0$-$S_1$ followed by an intersystem crossing to $T_1$ energy level, followed by a $T_1 \rightarrow T_2$ excited state absorption. RSA produced at this wavelength because of $T_1 \rightarrow T_2$ optical transition for is larger than that of $S_0 \rightarrow S_1$ transition. There might be a possibility of Two Photon Absorption (TPA) present in the samples. But, since TPA is highly intensity dependent, it was very less likely to be produced in the CW regime. The Peak – Valley configuration indicates the refractive nonlinearity is negative, i.e., self-defocusing. This is due to variation of thickness of the films with respect to solution temperature. The third-order nonlinear refractive index ($n_2$), the nonlinear absorption coefficient β, and the absolute value of $|\chi^{(3)}|$ was calculated from $|\chi^{(3)}| = [(\chi_R^{(3)})^2 + (\chi_I^{(3)})^2]^{1/2}$ for both the films where $\chi_R^{(3)}$ is the real and $\chi_I^{(3)}$ is the imaginary parts of third order nonlinear optical susceptibility. The calculated values from Z-scan curves are shown in the Table 3.

**TABLE 3.** Third order NLO parameters

| Sample | $n_2$ $10^{-8}$(m²/W) | β (m/W) | $\chi^{(3)}$ ($10^{-9}$) esu |
|---|---|---|---|
| 5 wt% Mn:BMZ | 1.8533 | 0.06561 | 1.22536 |

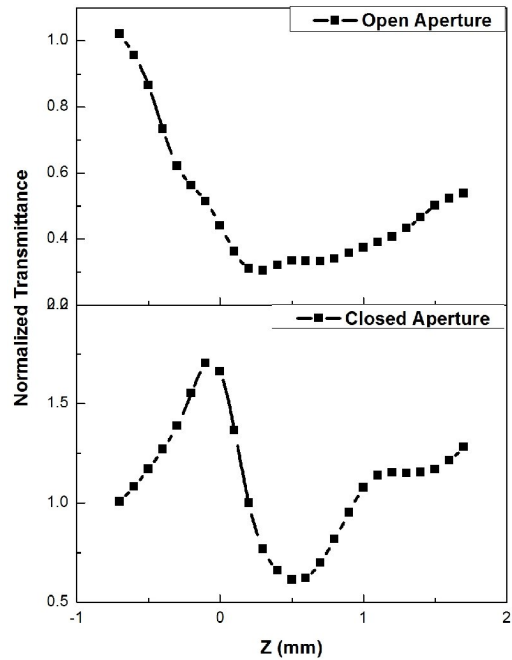

**FIGURE 3.** Open and Closed aperture curves of 5 wt% Mn doped BMZ Thin films.

## Conclusion

From the obtained results, it is concluded that the metal dopant in BMZ medium enhances its optical properties in optimum level. Also the deposition temperature plays a significant role in the quality of the deposited thin films.